# Population Vulnerability Models for Asteroid Impact Risk Assessment

Clemens M. Rumpf[1,*], Hugh G. Lewis[1], Peter M. Atkinson[2,3,4]

[1] University of Southampton, Engineering and the Environment, Southampton, UK

[2] Lancaster University, Faculty of Science and Technology, Lancaster, UK

[3] University of Southampton, Geography and Environment, Southampton, UK

[4] Queen's University Belfast, School of Geography, Archaeology and Palaeoecology, Belfast, UK

[*]Corresponding Author. Email  c.rumpf@soton.ac.uk

**Key Points:**

- Human vulnerability models for asteroid impacts for seven impact effects (e.g. tsunami, wind blast, etc)
- Introduction of risk and applicability to asteroid impact hazard.
- Protective function of continental shelf against tsunamis by impactors on shelf.

**Abstract**

An asteroid impact is a low probability event with potentially devastating consequences. The Asteroid Risk Mitigation Optimization and Research (ARMOR) software tool calculates whether a colliding asteroid experiences an airburst or surface impact and calculates effect severity as well as reach on the global map. To calculate the consequences of an impact in terms of loss of human life, new vulnerability models are derived that connect the severity of seven impact effects (strong winds, overpressure shockwave, thermal radiation, seismic shaking, ejecta deposition, cratering and tsunamis) with lethality to human populations. With the new vulnerability models ARMOR estimates casualties of an impact under consideration of the local population and geography. The presented algorithms and models are employed in two case studies to estimate total casualties as well as the damage contribution of each impact effect. The case studies highlight that aerothermal effects are most harmful except for deep water impacts, where tsunamis are the dominant hazard. Continental shelves serve a protective function against the tsunami hazard caused by impactors on the shelf. Furthermore, the calculation of impact consequences facilitates asteroid risk estimation to better characterize a given threat and the concept of risk as well as its applicability to the asteroid impact scenario are presented.

## INTRODUCTION

Earth has collided with asteroids since it was a planetesimal and this process continues albeit at a lower rate (Wetherill 1990); it is a natural phenomenon with potentially devastating consequences. Asteroid impacts have been responsible for at least two major disruptions in the evolution of life (Ryder 2002; Alvarez et al. 1980) and today, they remain a potential hazard for the human population (Popova et al. 2013; Chyba et al. 1993). Surveys scan the sky for asteroids in an effort to discover as many as possible and to calculate their orbits (National Research Council et al. 2010). Based on the propagation of orbits, the asteroids that may potentially impact the Earth in the future are identified subsequently. The European Space Agency (ESA) and the National Aeronautics and Space Administration (NASA), perform the collision detection using automated systems and the results, including impact probability and hazard rating, are published on their respective Near Earth Object (NEO) webpages (Universita Di Pisa and European Space Agency 2014; NASA 2014).

Two asteroid impact hazard scales are in use today: The Torino scale (Binzel 2000) aims to communicate the hazard rating of a given asteroid to the general public using an integer value from 0 (negligible hazard) to 10 (severe hazard) while the Palermo hazard scale (Chesley et al. 2002) is aimed at an expert audience using a continuous decimal numbering system where larger numbers indicate a higher hazard rating. Both scales have in common that they depend on the kinetic impact energy of the asteroid and its impact probability. In addition, the Palermo scale considers the time remaining until impact and compares it to the likelihood of a similarly energetic impact during that same timeframe. It is worth noting that both scales utilize kinetic impact energy as a proxy for the potential severity of an impact without specifically calculating the consequences of an impact. This observation is important because impact consequences are dependent on more parameters than only kinetic energy. These additional parameters include the impact angle, impact location (close to populated areas, in water, on ground), if the asteroid reaches the surface or explodes in mid-air, asteroid size, and the material characteristics of the asteroid and the surface. For example, an asteroid that enters the atmosphere at a shallow angle is more likely to experience an airburst than a surface impact with very different impact effects (e.g. lack of seismic shaking and cratering). Similarly, a large or dense asteroid is more likely to reach the surface than a small or highly porous one. Furthermore, the hazard of a large impact near a densely populated metropolitan area will yield more damage than a similar impact in an unpopulated desert. Therefore, it is crucial to consider the impact situation of each asteroid taking into account impact location as well as the physical circumstances (e.g. impact angle and speed) of the event.

At the University of Southampton, the Asteroid Risk Mitigation Optimization and Research (ARMOR) tool is under development to analyse the threat posed by discovered asteroids. ARMOR calculates impact effects and determines the lethally affected population in the impact zone considering their vulnerability. Consequently, ARMOR allows for the risk of known asteroids to be calculated in terms of expected casualties. This paper presents a new method for calculating asteroid risk including and emphasising the derivation of the necessary vulnerability models. The applicability of the vulnerability models is demonstrated by few examples that present estimation of total casualty numbers as well as the contribution of each impact effect in three scenarios.

## RISK AND VULNERABILITY MODELS

In this section, the concept of risk is defined, the percentage of the outside (of buildings) population is derived and vulnerability models are presented.

**Risk**

The concept of risk is applicable to a wide variety of subjects (e.g. finance, insurance, politics and decision making). Risk, defined as the expected loss, is the product of three factors: the probability that an event occurs, exposure, the value that is at stake (or exposed), and vulnerability - the portion of the exposed value that is affected if the event occurs. Specifically, for the asteroid impact hazard, this relation can be stated in mathematical terms as:

$$R = P \times \psi \times V(S) \qquad (1)$$

where $R$ is asteroid hazard risk, $P$ is the asteroid impact probability, $\psi$ is the population (exposure) and $V(S)$ is the vulnerability which is a function of the severity $S$ of harmful effects generated by an asteroid impact. The information needed to assess the asteroid impact probability is provided by ARMOR as the asteroid's spatially distributed impact probability (shown in (Rumpf et al. 2016b)) which not only allows identification of the possible impact locations but also provides information about localized impact probability. The global population map feeds the exposure term and provides exposure values as well as its spatial distribution. Here, the global population map for the year 2015 (CIESIN et al. 2005) with a grid resolution of 4.6×4.6 km$^2$ (Figure 1) is used. Vulnerability describes what portion of the exposed population is lethally affected by the asteroid impact and this term depends on the severity of the impact generated effects. The process of impact effect modelling and vulnerability estimation is described in the following sections along with the derivation of vulnerability models.

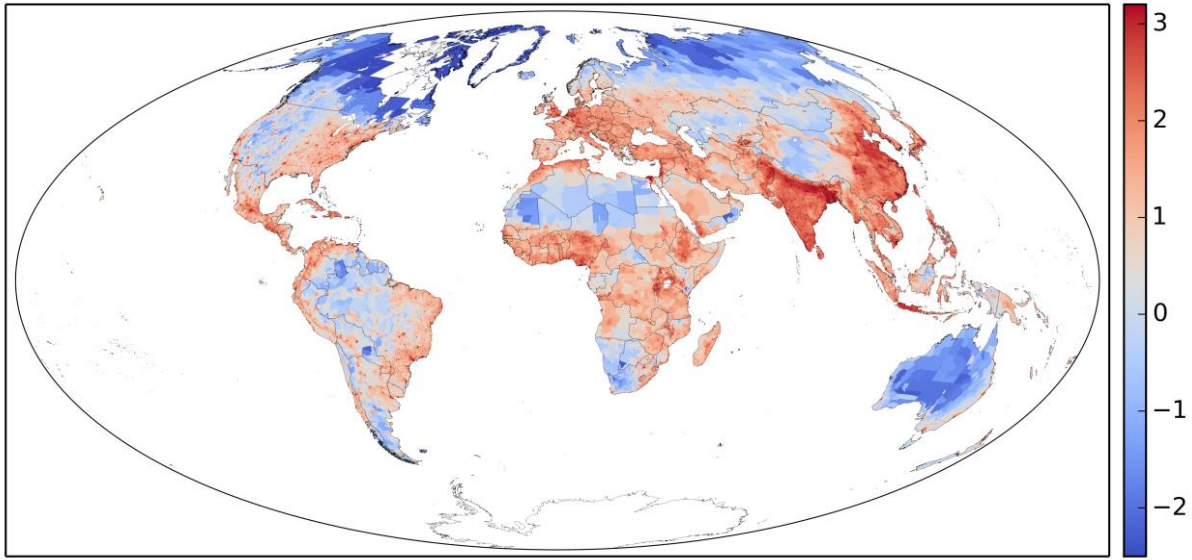

Figure 1: World population density map for the year 2015 based on data by (CIESIN et al. 2005). Data represents population density as people per square km and the scale represents powers of ten. Note that maximum population density is higher than represented by this scale in certain regions.

**Unsheltered Population**

For subsequent vulnerability analysis, it was necessary to define the average percentage of global population that is unsheltered. Unsheltered population was defined as any population that is outside of buildings and is, thus, more susceptible to environmental effects.

The literature provides some data about the average time that people spend outdoors but the used datasets are limited to populations that share similar work patterns with the so called "western world". Reference (Klepeis et al. 2001) finds that the average American spends 13% ≈ 3.12 hours per day outside buildings and the meta study (Diffey 2011) reports that people belonging to western nations spend an average of 1.99 hours per day outdoors which does not include time spent in vehicles. Vehicles offer negligible shelter against thermal radiation as well as shock waves and the time spent in vehicles was counted towards unsheltered time where commuting time was used as a proxy for time in vehicles. The Labour Force (Trades Union Congress 2012) reports that the average commuting time in the UK in 2012 was 54.6 minutes. Similarly, the U.S. Census American Community Survey (U.S. Census Bureau 2011) indicates that the average round-trip commuting time in the United States is 50.8 minutes. Adding commuting time as well as the outside 1.99 hours from the meta study provides the time spent outdoors as supported by the meta study and this time is about 2.87 hours or about 12% of each day. Together, the findings indicate that the average westerner is unsheltered for about 13% of each day.

The population that the above work pattern was applied to is about 2.5 billion people (European Union, USA, Canada, Australia, New Zealand, Japan and parts of: Russia, China, India, Brazil, Argentina, Arab countries), while the global population is about 7.3 billion people. The data reported above do not account for non-western populations and given the lower industrialisation standard in non-western countries, it is assumed that non-western populations spend twice as long outside as westerners (26%). With this assumption the western and non-western populations could be connected and the weighted average time that the global population spends outdoors was computed to 22% per day after:

$$\frac{0.13 \times 2.5 + 0.26 \times (7.3 - 2.5)}{7.3} = 0.22 \quad (2)$$

For further analysis, it was assumed that 22% of the global population is unsheltered at any given time.

**Impact effect and vulnerability modelling**

Upon colliding with the Earth, an asteroid deposits most of its energy either in the atmosphere, during an airburst, or on the surface after it passes the atmosphere mostly intact. Whether a surface impact or airburst occurs depends on the entry conditions of the asteroid: impact angle, impact speed, size of the asteroid, and material. In this analysis, impact angle and speed are provided by ARMOR's orbit dynamic impact simulation. Furthermore, size values are published by ESA and NASA on their NEO webpages and the sizes were estimated based on the asteroid's brightness. Finally, the asteroid body was assumed to be similar to ordinary chondrites with a density of 3100 $kg/m^3$ corresponding to an estimated yield strength of 381315 Pascal (Pa) (Collins et al. 2005). Ordinary Chondrites account for about 90% of all known meteorites (Britt 2014).

The process that was used to determine if an asteroid experiences an airburst or surface impact is visualized in Figure 2 and it is based on (Collins et al. 2005). This process employs analytical models to calculate the outcomes of physical processes that occur during atmospheric passage (e.g. break-up altitude, airburst altitude, impact velocity, etc.) as well as the severity of subsequent impact effects. Once the asteroid airbursts or impacts the surface, its energy is released in a variety of impact effects and in this analysis, seven impact effects are modelled: High winds, overpressure, thermal radiation, cratering, seismic shaking, ejecta blanket deposition and tsunami. The first three of these may occur in both, airburst or surface impact, while the latter four occur only in a surface impact.

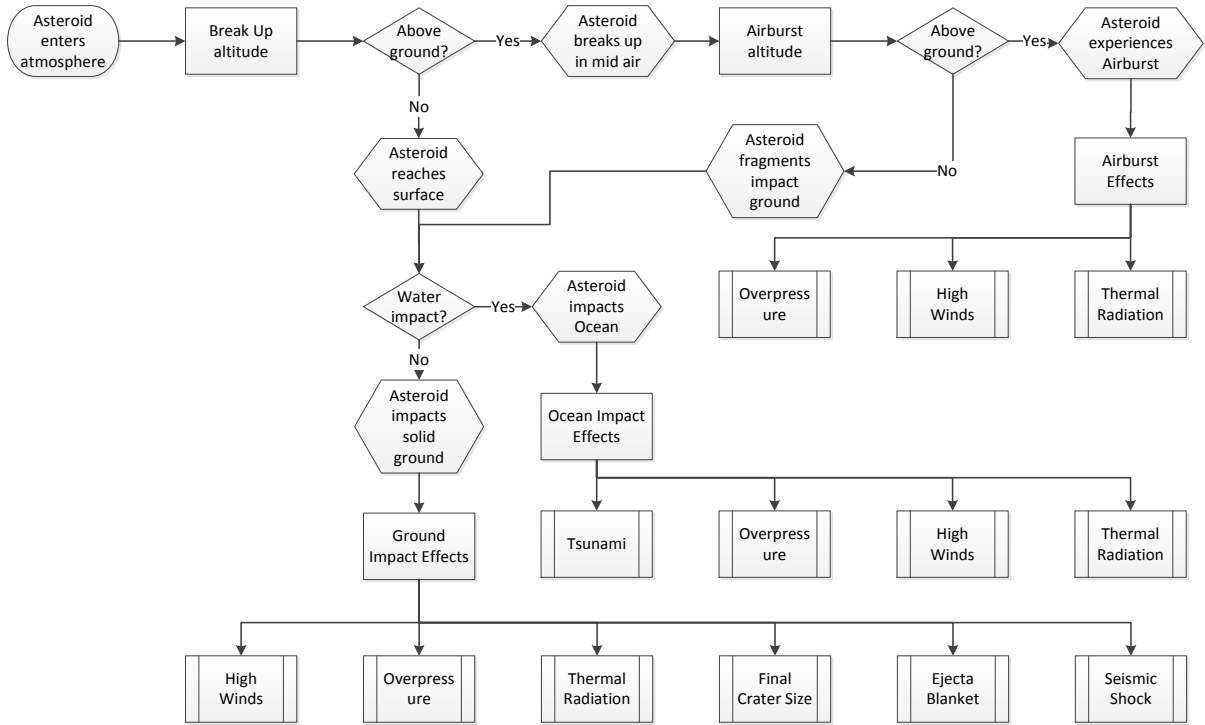

Figure 2: Impact effect flow diagram showing how an airburst or surface impact is determined and the corresponding impact effects.

Upon airburst or impact, the asteroid's kinetic energy is released in the form of impact effects and these impact effects are greatest at the impact site. Starting from the impact site, the effects propagate outwards and attenuate with greater distance. The strength of an effect is called severity and the more severe an effect is, the more likely it is that the population is harmed. In other words, higher severity yields increased population vulnerability.

The following sections describe all seven impact effects and their effects on the population. Most effect models are described in greater detail in (Collins et al. 2005) while tsunami modelling required detailed treatment here. Vulnerability models were not readily available, and most vulnerability models presented here were the result of a combinatory literature review of partially available models coupled with evidence based model derivations to fill gaps in the literature. A notable resource for vulnerability research is (Glasstone and Dolan 1977) and other sources are indicated where applicable. Vulnerability models are usually represented by sigmoid functions of the form:

$$V_{effect}^{case}(S) = a \frac{1}{1 + e^{b(S+c)}} \qquad (3)$$

where $V_{effect}$ is the vulnerability to a given impact effect, $S$ is the severity of the effect, and $a$, $b$ and $c$ are constants that are determined in the following sections.

After the vulnerability of each impact effect is estimated based on effect severity, the combined vulnerability is calculated which provides the portion of the exposed population that adds to the casualty count. An efficient way to calculate combined vulnerability is to determine the chance that an individual survives all impact effects. Vulnerability may also be understood as the likelihood that an individual will die due to impact effects and this concept helps to determine the combined vulnerability. Since vulnerability $V_{effect}$ is the chance that an individual dies through one of the impact effects, the term $\lambda_{effect} = 1 - V_{effect}$, conversely, describes the chance that this individual survives the impact effect. After all effect vulnerabilities, and, similarly, all effect survivability chances, are computed, the chance of an individual to survive all effects in sequence may be calculated as:

$$\lambda_{combined} = \prod_{i=effects} \lambda_i \quad (4)$$

Finally, the combined vulnerability of all impact effects is: $V_{combined} = 1 - \lambda_{combined}$.

**High Winds and Overpressure**

During an airburst or impact. the asteroid deposits its energy in an explosion like event that produces an aerodynamic shockwave resulting in a tornado like wind gust and overpressure peak. In accordance with (Collins et al. 2005), overpressure in a ground impact is calculated as:

$$p_D = \frac{p_x D_x E_{kt}^{1/3}}{4}\left(1 + 3\left[\frac{D_x E_{kt}^{1/3}}{D}\right]^{1.3}\right) \quad (5)$$

where $p_D$ is pressure in Pa at distance $D$ from the impact point in meters, $p_x = 75000$ Pa and $D_x = 290$ m are scaling parameters and $E_{kt}$ is the asteroid's kinetic energy at the time of energy deposition in equivalent kilo tons of Trinitrotoluene (kt TNT). In an airburst event, the overpressure shockwave reflects off the surface of the Earth. Directly below the airburst point, a simple shockwave arrives at the surface and overpressure [Pa] is (Collins et al. 2016):

$$p_D = p_0 e^{-\beta D\left(E_{kt}^{-1/3}\right)} \quad (6)$$

where

$$p_0 = 3.14 \times 10^{11} z_{b1}^{-2.6} + 1.8 \times 10^7 z_{b1}^{-1.13} \quad (7)$$

$$\beta = 34.87 z_{b1}^{-1.73} \quad (8)$$

The calculation of energy scaled airburst altitude $z_{b1}$ is described in (Collins et al. 2005) as scaling the result of equation (18) using equation (57) of that reference. The pressure shockwave is reflected from the surface and it interacts constructively with the original shockwave at sufficient distance from the airburst. In fact, this condition is already described in equation (5) and the switching distance $D_{m1}$ in meters between equations (6) and (5) is:

$$D_{m1} = \frac{550 \, z_{b1}}{1.2(550 - z_{b1})} \quad (9)$$

High winds realized in tornado-like wind gusts are a result of the overpressure shockwave and the wind speed $v_{wind}$ in meters per second, after (Glasstone and Dolan 1977), is:

$$v_{wind} = \frac{5 p_D}{7 p_a} \frac{c_0}{\left(1 + {6 p_D}/{7 p_a}\right)^{0.5}} \quad (10)$$

where $p_a$ is the ambient pressure and $c_0$ is the speed of sound.

The overpressure $p_D$ and wind speed $v_{wind}$ describe the severity of these two impact effects. Effect severity was used to determine the vulnerability of the populations that live within the area that is affected by an impact.

Overpressure injures humans by creating a harmful pressure differential between the organ internal pressure (lungs) and ambient pressure. The shockwave rapidly increases ambient pressure leaving the body internals insufficient time to adjust and the resulting pressure differential can rupture tissue.

For overpressure vulnerability, three sigmoid functions were fitted to experimental data presented in (Glasstone and Dolan 1977) (Table 12.38). In addition to an expected vulnerability model $V_p^{expect}$, that uses the median values in the table, best $V_p^{best}$ and worst $V_p^{worst}$ case vulnerability functions were derived based on the value ranges provided in the table. The purpose of adding worst and best cases is to gain a sense of the sensitivity of the impact effect models. The resulting overpressure vulnerability $V_p$ models are dependent on overpressure $p_D$ (at a given distance) and the best fit values for the coefficients $a$, $b$ and $c$ for the best, expected and worst cases are:

Table 1: Overpressure vulnerability coefficients.

| Case | $a$ | $b$ | $c$ |
|---|---|---|---|
| Expected | 1.0 | $-2.424 \times 10^{-5}$ | $-4.404 \times 10^5$ |
| Best | 1.0 | $-1.899 \times 10^{-5}$ | $-5.428 \times 10^5$ |
| Worst | 1.0 | $-2.847 \times 10^{-5}$ | $-3.529 \times 10^5$ |

Thus, the vulnerability function to overpressure is:

$$V_p^{case}(p_D) = a\ \frac{1}{1 + e^{b(p_D+c)}} \quad (11)$$

The vulnerability function is plotted in Figure 3 along with the experimental data points.

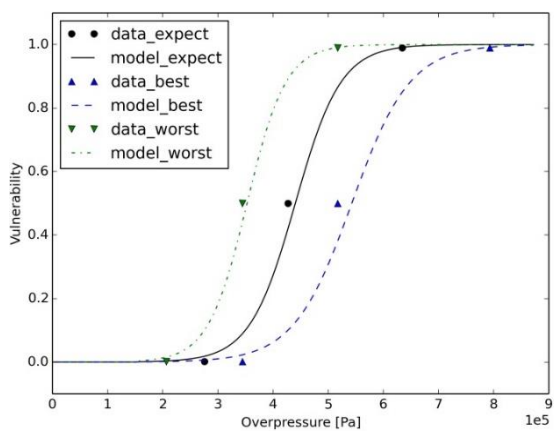

Figure 3: Overpressure vulnerability models with experimental data points.

In addition to causing direct injuries to humans, an overpressure shock may cause buildings to collapse resulting in further fatalities. This point should be considered when interpreting results but is not included in the present vulnerability model.

Strong winds accompany the overpressure shockwave and the severity of strong winds is expressed by equation (10). In fact, overpressure shockwave and strong winds occur together and depend on each other. However, they are treated separately in terms of vulnerability models because their mechanism of harming humans differs (overpressure: internal organs, wind: dislocation of bodies or objects). A wind vulnerability model was derived based on the severity making use of the similarity between strong wind gusts and the criteria in the Enhanced Fujita (EF) scale, which is used to classify tornado strength (Wind Science and Engineering Center 2006) in the United States of America (USA). In the EF scale, tornados are classified based on the damage that they cause during the peak 3 seconds wind gust and Table 2 provides an overview of the EF category, wind speed and expected damage.

Table 2: Enhanced Fujita scale.
Categories, wind speeds and damage

| Category | 3s Wind-Gust [m/s] | Typical Damage |
|---|---|---|
| EF0 | 29-38 | Large tree branches broken; Trees may be uprooted; Strip mall roofs begin to uplift. |
| EF1 | 38 - 49 | Tree trunks snap; Windows in Institutional buildings break; Facade begins to tear off. |
| EF2 | 49 - 60 | Trees debark; Wooden transmission line towers break; Family residence buildings severely damaged and shift off foundation. |
| EF3 | 60 - 74 | Metal truss transmission towers collapse; Outside and most inside walls of family residence buildings collapse. |
| EF4 | 74 - 89 | Severe damage to institutional building structures; All family residence walls collapse. |
| EF5 | >89 | Severe general destruction. |

*EF0*

According to the EF scale, EF0 corresponds to wind speeds between 29-38 m/s. Humans can be harmed in this condition by being thrown against objects or objects being hurled at them. In (Glasstone and Dolan 1977) lethality estimates are provided for objects turned missiles that hit the body. According to this source, a 5 kg object entails a near 100% rate of fracturing a skull when hitting the head with a velocity exceeding 7 m/s. Furthermore, lethality may occur when the body is thrown against solid objects with velocities in excess of 6 m/s. It is conceivable that these events may be produced in a category EF0 tornado and, indeed, category EF0 tornados have been lethal in the past (NOAA 2015) but the casualty rate is low (3 people were killed by EF0 tornados between 1997 and 2005). Here, it was assumed that 1% of the population that is outdoors is hit by missiles or thrown against objects (affected population) and that 2% of these individuals die as a direct result of the injury. These assumptions provide a vulnerability of 0.000044 for strong winds corresponding to a category EF0 tornado.

*EF1*

With increasing wind speeds, a larger portion of the outside population will be affected; more people will be thrown against solid objects because the strong wind will be able to lift up more people. The wind will also generate more missiles that could hit victims. Furthermore, the lethality for each person also increases because the impact speed of the body or the missile will be higher. Reference (Glasstone and Dolan 1977) estimates that 50% lethality is reached when a body contacts a solid object with a speed of 16.5 m/s and 100% lethality is reached at 42 m/s. It seems plausible that a body could be accelerated to speeds of 16.5 - 42 m/s in an EF1 tornado. However, it can be assumed that some of the outside population finds sufficient shelter. Hence, it was assumed that 10% of the outside population is affected and that 5% of those affected die. Housing still provides good protection against EF1 level winds but it was assumed that 1% of the inside population can be affected and that 5% of those affected die. Vulnerability for winds corresponding to an EF1 tornado was, thus, set to 0.0015.

The above assumed increase in vulnerability agrees well with the increase of lethality of recorded tornados between 2000 and 2004 (NOAA 2015). During that time period 4284 EF0 tornados killed 2 people resulting in a casualty rate of 0.00047 per EF0 tornado. In the same time 1633 EF1 tornados killed 20 people yielding a casualty rate of 0.012 per EF1 tornado which is a 26 fold increase. Similarly, assumed vulnerability for strong winds increased by a factor of 33.

*EF2*

Increasing wind speed renders shelters less effective as houses start to exhibit significant damage. It is assumed that, in addition to 40% of the outside population, 5% of the housed population is affected yielding a total of 12.7% affected population. Lethality for the affected population increases to 10% as wind speeds are capable of accelerating bodies beyond the 42 m/s body impact speed assumed for 100% lethality (Glasstone and Dolan 1977) and objects turned missiles have higher damage potential. Consequently, vulnerability is equal to 0.013.

The increase in vulnerability from EF1 to EF2-like wind speeds of a factor of 8.5 matches the casualty rate increase from EF1 to EF2 tornados. Between 2000 and 2004, 439 EF2 tornados killed 51 persons yielding a casualty rate of 0.12 per EF2 tornado corresponding to a 9.6 fold increase.

*EF3*

Tornados of this category destroy most housing shelter leaving basements and well-constructed concrete buildings as viable shelter options. It was assumed that 80% of the outside and 30% of the inside population would be affected by winds of this strength. For those affected outside and inside, lethality increased to 30% and 20%, respectively, due to hitting missiles or by being thrown against fixed structures. The vulnerability thus increases 8 fold to 0.10. The record shows that 116 persons were killed by 127 EF3 tornados yielding a casualty rate of 0.913 that corresponds to an 8 fold increase from EF2 to EF3 tornados.

In fact, (Paul and Stimers 2014) show that the vulnerability inside a zone affected by an average of EF3 tornado winds (Kuligowski et al. 2013), was 2.1%. In contrast to a tornado, the windblast in the case of an unforeseen asteroid impact would arrive without prior warning by the government or by meteorological cues that the population could be expected to correctly interpret. It is shown in (Simmons and Sutter 2005) and (French et al. 1983) that the presence of a warning decreases mortality by a factor of about three. In addition, housing standards in the USA ensure that protection of the population against windblast is better than the global average by an assumed 50% (factor 1.5). Taking into account the influence of warning and better protection, the observed vulnerability of 2.5% can be expected to increase to a global average of over 9% matching well with the windblast vulnerability found here.

*EF4*

Persons who are sheltered in very well constructed concrete buildings will be protected against these winds. It was assumed that 40% of the inside and 90% of the outside population would be affected with corresponding lethality rates of 30% and 40%, respectively. Thus, vulnerability is 0.17.

*EF5*

The great majority of structures collapse in these winds offering diminishing protection. Consequently, it is assumed that 95% of the outside and 50% of the inside population is affected with a lethality rate of 50% and 40%, respectively. The resulting population vulnerability is 0.26 at 89 m/s wind speed.

Evidently, (Wurman et al. 2007) modelled EF5 tornados in an urban setting and assumed that 10% of the inside population would be affected lethally. Taking into account the criticism that this is likely an overestimation for a setting in the USA, 5% lethality seems more likely. Considering the influence of warning (factor 3) and extrapolating to a global setting (factor 1.5), a value of 22.5% was obtained which correlates closely to the value found previously.

Based on these data, three vulnerability models were derived: One model that describes the expected case $V_{wind}^{expect}$ and two for a worst $V_{wind}^{worst}$ and best $V_{wind}^{best}$ case. The expected case uses the median wind speed for each EF category with the corresponding vulnerability value, while the worst and best case models use the wind speeds of one category lower or higher, respectively. The model function is:

$$V_{wind}^{case}(v_{wind}) = a \frac{1}{1 + e^{b(v_{wind}+c)}} \quad (12)$$

and the corresponding coefficients are:

Table 3: Wind vulnerability coefficients.

| Case | a | b | c |
|---|---|---|---|
| Expected | 1.0 | $-5.483 \times 10^{-2}$ | $-1.124 \times 10^{2}$ |
| Best | 1.0 | $-5.036 \times 10^{-2}$ | $-1.293 \times 10^{2}$ |
| Worst | 1.0 | $-5.549 \times 10^{-2}$ | $-9.898 \times 10^{1}$ |

Figure 4 shows the vulnerability models plotted over the relevant range of wind gust speeds.

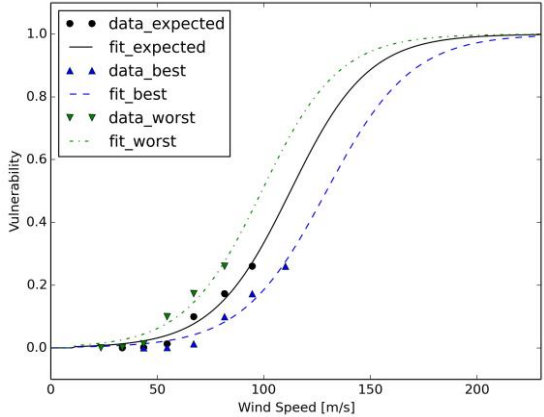

Figure 4: Wind vulnerability models with data points.

**Thermal Radiation**

Surface impacts as well as airburst produce thermal radiation but the two events require separate modelling as presented in the following.

*Ground Impact*

If the impacting meteoroid travels in excess of 15 km/s, enough energy is released to evaporate the asteroid and some of the ground material. This violent event generates a plume with very high pressure (>100GPa) and temperature (≈10000K) that rapidly expands. This is called the fireball. As a result of the high temperature, the gas is ionized and appears opaque to thermal radiation due to the plasma's radiation absorption characteristics. Consequently, the plume expands adiabatically and only starts to radiate outwards when the plasma cools to the transparency temperature $T_*$ (Zel'dovich and Raizer 1966). (Collins et al. 2005) report an empirical relationship for the fireball radius $R_f$ when it reaches transparency temperature as a function of impact energy $E$:

$$R_f = 0.002 E^{1/3} \quad (13)$$

Only a fraction of the kinetic energy released during impact is transformed into thermal radiation (Nemtchinov et al. 1998). This fraction is called the luminous efficiency $\eta_{lum}$ and (Ortiz et al. 2000) determined that it is on the order of $10^{-4}$ to $10^{-2}$. The received thermal energy per area unit (assuming a hemispheric dissipation of heat radiation) is given by (Collins et al. 2005) as:

$$\phi = f \frac{\eta_{lum} E}{2\pi D^2} \quad (14)$$

where, $f$ is the fraction of the fireball that is visible over the horizon at distance $D$ which is also a function of $R_f$ and the corresponding geometric relationship is given in (Collins et al. 2005) (equation (36)). The effect severity of thermal radiation is subject to additional considerations such as local weather (e.g. fog will reduce severity) and the topography providing shadowing opportunities. Additionally, luminous efficiency is currently not well constrained and these considerations require a cautionary note that thermal radiation severity could vary significantly from estimates provided by the present model. This comment is also valid for the severity estimation of other impact effects as uncertainty in severity estimation (Collins et al. 2005) is not considered in the best and worst case scenarios here which are designed to capture uncertainty in vulnerability modelling only,

**Airburst**

Besides the air blast, some of the kinetic energy carried by the meteoroid that is released during airburst dissipates as thermal radiation. (Nemtchinov et al. 1994) investigates the radiation emitted by meteors and the following airburst thermal radiation model was derived here based on this research. Equation (11) of the reference provides an expression for thermal energy flux density based on airburst intensity:

$$\phi = q_h \left[\frac{L_0}{D_{los}}\right]^2 5 \qquad (15)$$

where $\phi$ is the energy flux density in [W/m$^2$] at the target distance, $q_h$ (the reference uses $q_\infty$) is the energy flux density of the meteoroid at a given altitude, $L_0$ is the asteroid diameter and $D_{los}$ is the line of sight distance from the airburst to the target. Table 1 of the reference provides values for $q_h$ as a function of speed for the two altitudes of 25 km and 40 km. Here, an interpolation function was built that produces $q_h$ values for any given speed, altitude pair based on table 1 in the reference. To this end, a six degree polynomial was least square fitted to the data describing $q_{h=25}$ at 25 km altitude as a function of meteoroid speed $v$:

$$\begin{aligned}q_{h=25} = &(-4 \times 10^{-16}\, v^6) \\ &+ (7 \times 10^{-11}\, v^5) - (5 \times 10^{-6}\, v^4) \\ &+ (0.176\, v^3) - (3160.6\, v^2) \\ &+ (3 \times 10^7\, v) - 1 \times 10^{11}\end{aligned} \qquad (16)$$

The polynomial has a correlation coefficient of 0.9868 with the data. Only three data points were available for the data for $q_{h=40}$. However, the data are fitted perfectly by the line described by:

$$q_{h=40} = 700000\, v - 1 \times 10^{10} \qquad (17)$$

Finally, a linear interpolation scheme estimates $q_h$ for any given airburst altitude $z_b$ based on the calculated values for $q_{h=25}$ and $q_{h=40}$.

$$q_{h=z_b} = \frac{q_{25} - q_{40}}{15000}(40000 - z_b) + q_{40} \qquad (18)$$

The distance $D_{los}$ was estimated using Pythagoras' relationship with airburst altitude $z_b$ and surface distance $D$ as parameters:

$$D_{los} = \sqrt{z_b^2 + D^2} \qquad (19)$$

With these relations, equation (15) can be solved and a thermal energy flux density may be obtained for any airburst event.

Note that the unit of equation (15) is [W/m$^2$] and that for subsequent analysis the thermal radiation energy density [J/m$^2$] was needed. Based on visual observations of the Chelyabinsk (Popova et al. 2013) and other meteors (Borovička and Kalenda 2003) it was determined that a break-up occurs within a time span on the order of one second. Therefore, one second was assumed as the default break-up duration for airbursts and the unit [W/m$^2$] is equivalent to the energy density [J/m$^2$] when integrated for this timespan because energy is the integral of energy flux [W]=[J/s] over time [s]. This relation is expressed by the following example equation assuming that energy flux is constant over time:

$$1\text{J} = 1\text{W} \times 1\text{s} = 1\text{ J/s} \times 1\text{s} \qquad (20)$$

*Vulnerability Model*

Thermal radiation is emitted from airbursts and surface impacts. Surfaces that are incident to the radiation heat up and can be scorched or ignited. The consequences of thermal radiation energy exposure on the human body as a consequence of nuclear detonations were investigated in (Glasstone and Dolan 1977) and this serves as the basis for the thermal radiation vulnerability model. It should be noted that the spectral intensities in the burn relevant portion of asteroid and nuclear explosion generated radiation spectra will differ from each other. This could lead to non-identical efficiencies in translating radiation energy into burn injury. However, given the sparse evidence basis of asteroid explosions and few literature sources, the approach presented here represents a best effort to treat asteroid caused radiation vulnerability. The burn probability as a function of radiant exposure and explosion yield is given in Figure 12.65 of the reference. While the dependency of burn probability to radiant exposure [$J/m^2$] is obvious, its dependency on explosion yield should be explained.

The dependency on explosion yield is rooted in the observation that the process of small yield explosions takes less time to unfold than large yield explosions resulting in different energy flux rates. For smaller explosions, a given amount of radiant energy is delivered in a shorter time compared to a larger explosion and, thus, the radiation intensity differs with explosion yield. Higher radiation intensity causes injuries more readily than low intensity radiation even though the same cumulative energy might be delivered in both cases. The reason for this behaviour is that the heated surface has more time to dissipate the incident radiation energy in a low intensity radiation case. Unlike nuclear explosive devices, meteoroids are not optimized for explosion and it is thus assumed that their explosion signature is more comparable to that of a large nuclear device because the explosion process takes a relatively long time. The data used to build the vulnerability model correspond to the results produced by a 1Mton TNT equivalent yield nuclear device as shown in Figure 12.65 in (Glasstone and Dolan 1977).

The burn severity distribution is a function of radiant exposure and the data in the reference forms the basis for Figure 5 which shows the burn degree that can be expected when exposed to a certain radiant energy.

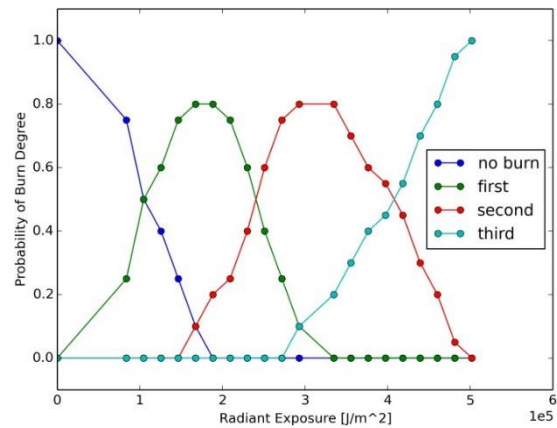

Figure 5: Burn degree distribution as a function of radiation intensity based on data in (Glasstone and Dolan 1977) assuming the explosion signature of a 1Mton TNT yield nuclear device.

Aside from burn degree, the total body surface area (TBSA) of a human that is burned determines the expected mortality. In (American Burn Association 2012), statistical analysis of 143199 burn victims in the United States were analysed for their mortality rate based on burned TBSA. The reported numbers apply to persons who have been treated in medical facilities after the burn injury. This means that the burn injury itself could be treated adequately but also that possible subsequent medical complications (pneumonia, infection) that are linked directly to the burn injury could be addressed. Here, it shall be assumed that mortality rates are twice as high because proper and timely treatment of burn injuries is unlikely in the event of an asteroid impact that will potentially affect a large region and its medical infrastructure. Figure 6 visualizes the data in Table 9 of (American Burn Association 2012) and shows the mortality rate as a function of burned TBSA for treated and untreated victims.

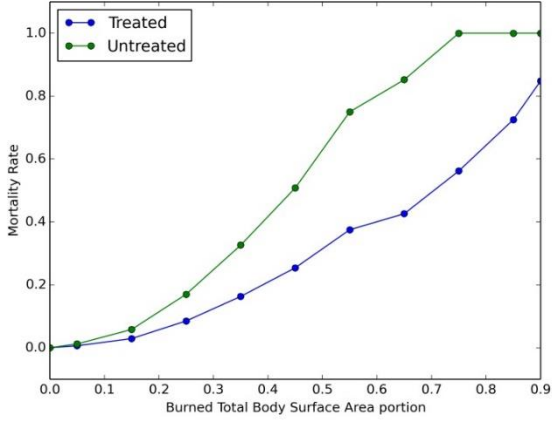

Figure 6: Mortality rate for treated and untreated burn victims as a function of burned TBSA. Data from (American Burn Association 2012).

To relate radiant exposure to TBSA and, thus, to mortality rate, a scaling law is introduced that approximates TBSA based on the burn degree distribution as a function of radiant exposure. In general, every part of the body that is exposed to light from the meteoroid explosion will be burned, but the severity of the burn differs. A superficial first degree burn, which is comparable to bad sunburn, is less life threatening than a third degree burn that penetrates through all skin layers. To account for this distinction, a scaling law was introduced that yields TBSA as a function of burn degree distribution. The scaling law is the weighted sum (first degree has weight one, second degree has weight two and third degree has weight three) of the burn distribution as a function of radiant exposure.

$$TBSA_{weighted}(\phi) = \frac{1}{9}[1 \times burn_{1°}(\phi) + 2 \times burn_{2°}(\phi) + 3 \times burn_{3°}(\phi)] \quad (21)$$

Furthermore, the scaling law respects the observation that the thermal radiation from an asteroid impact arrives from only one direction. This situation suggests that only half of a human, or a maximum of 50% TBSA, can be injured from thermal radiation. Moreover, clothing (as long as it does not burn itself) provides protection against a short lived energy burst of thermal radiation and it is therefore assumed that only one third of TBSA can be burned for people standing outside. Figure 7 visualizes the resulting TBSA curve as a function of radiant exposure.

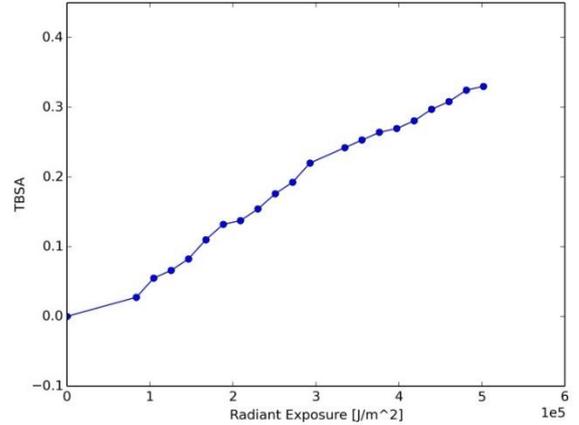

Figure 7: Visualization of TBSA-burn degree scaling law (equation (21)). The maximum TBSA is scaled to one third as clothing offers protection and radiation comes from one direction.

Combining the data from (Glasstone and Dolan 1977) about radiant exposure and the resulting burn severity with the scaling law to relate burn severity with TBSA and, finally, with the data from (American Burn Association 2012) about mortality rate based on TBSA, mortality rate can be expressed as a function of radiant exposure. Figure 8 shows the relationship. The data are based on recorded occurrences and the corresponding radiant exposure range is limited to these records. An asteroid impact can produce higher radiant energies and the mortality rate, thus, has to be expanded to larger values of radiant exposure.

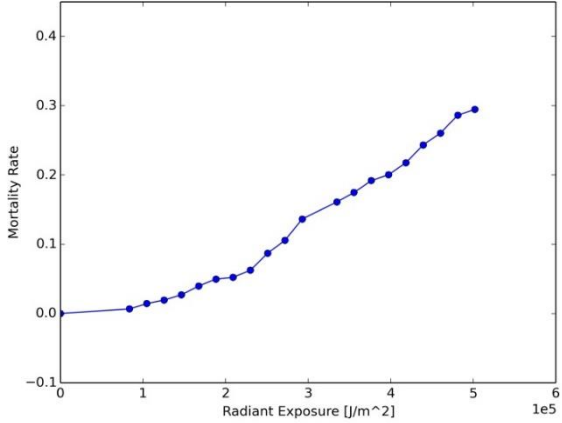

Figure 8: Mortality rate as a function of radiant exposure.

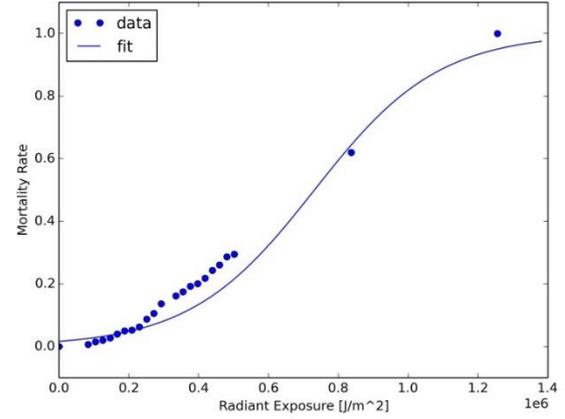

Figure 9: Mortality rate as a function of the full, applicable radiant exposure range.

Clothing provides limited protection to thermal radiation because it can absorb thermal energy up to the point when it itself ignites. It was therefore assumed that only one third of TBSA can be burned before clothing ignites. Reference (Glasstone and Dolan 1977) reports that cotton and denim clothing ignites at about 836800 J/m$^2$. Beyond this energy level clothing does not offer protection and it was assumed that 50% of TBSA can be burned resulting in a mortality rate of 62% (American Burn Association 2012). Furthermore, at energy densities of 1255200 J/m$^2$, (Glasstone and Dolan 1977) reports that sand explodes (popcorning), aluminium aircraft skin blisters and roll roofing material ignites. These conditions appear lethal to humans and a mortality rate of one is assumed for a population exposed to this energy level.

Figure 9 presents the full range of thermal radiation mortality rate and shows the corresponding data points. Additionally, a sigmoid function has been least square fitted to the data as follows:

$$Mortality\ Rate(\phi) = \frac{1}{1 + e^{-0.00000562327(\phi - 731641.664)}} \quad (22)$$

The mortality numbers derived above apply to the exposed population that is outside sheltering buildings. For people inside buildings the mortality rate will be moderated through the protective effect of walls. However, windows do not offer protection against thermal radiation and it is assumed that one third of the inside population (25% of global population) is exposed through windows even though they are inside a building. The expected case is that 22% are outside and 25% are exposed behind windows (totalling 47%) while the remaining 53% of the global population are unaffected by thermal radiation. The mathematical expression for thermal radiation vulnerability is, thus:

$$V_\phi^{case}(\phi) = a\ \frac{1}{1 + e^{b(\phi+c)}} \quad (23)$$

For the expected case, a maximum of 47% of the population is exposed. Additionally, in the worst case scenario it was assumed that the entire population is outdoors (exposed) while in the best case scenario the entire population is sheltered and the corresponding coefficients are:

Table 4: Thermal radiation vulnerability coefficients.

| Case | a | b | c |
|---|---|---|---|
| Expected | 0.47 | $-5.623 \times 10^{-6}$ | $-7.316 \times 10^5$ |
| Best | 0.25 | $-5.623 \times 10^{-6}$ | $-7.316 \times 10^5$ |
| Worst | 1.0 | $-5.623 \times 10^{-6}$ | $-7.316 \times 10^5$ |

Figure 10 visualizes these vulnerability models.

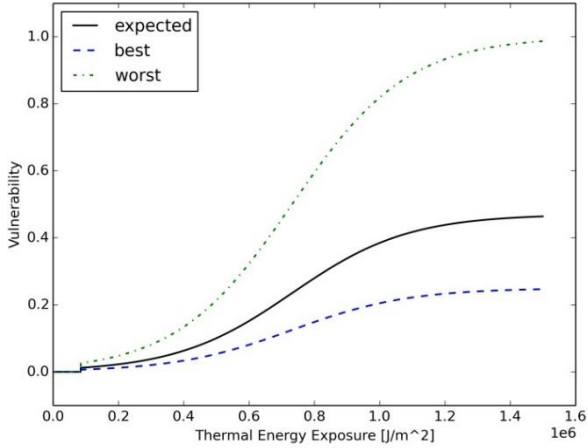

Figure 10: Thermal radiation vulnerability models as functions of thermal energy exposure.

**Cratering**

When a meteoroid impacts the surface, an impact crater forms. The cratering process is complex in itself and occurs in several steps. In a first step, a transient crater is formed which is the dynamical response to the impacting meteoroid. It is useful to calculate the transient crater because the final crater shape depends on the intermediate step of the transient crater. In fact, the energy delivered by the asteroid is so large and the speed of the mechanical interaction between asteroid and surface is so fast that the target material (water or ground) react like a fluid and thus can be described with the same formalism. A transient crater is generally an unstable structure and is similar to the crown-like shape that forms in a water surface immediately after a droplet falls into it. The "crown ring" surrounds the impact point that forms a bowl shaped depression and represents the crater bottom. A transient crater is not self-supporting and collapses under the influence of gravity to form the final crater shape. The transient crater diameter $D_{tc}$ is given in (Collins et al. 2005) with:

$$D_{tc} = 1.161 \left(\frac{\rho_i}{\rho_t}\right)^{1/3} L_0^{0.78} v_i^{0.44} g_0^{-0.22} \sin^{1/3}\gamma \quad (24)$$

where $\rho_i$ is the impactor density, $\rho_t$ is target (ground) density (assumed to be 2500 kg/m$^3$), $v_i$ is impactor speed, $g_0 = 9.80665$ m/s$^2$ is Earth standard gravity and $\gamma$ is the impactor angle (an impact velocity vector normal to the surface corresponds to $\gamma = 90°$).

With the collapse of the transient crater, the final crater forms. For simple craters, up to 3.2 km in diameter on Earth, final crater diameter is linearly related to the transient crater diameter according to Equation (22) in (Collins et al. 2005):

$$D_{fr} = 1.25 D_{tc} \quad (25)$$

For complex craters, final crater scaling is a nonlinear function of transient crater size (Equation 27 in (Collins et al. 2005)); however, since the focus here was asteroids up to 500-m in diameter, which form craters only just above the simple-to-complex transition, simple crater scaling for all crater sizes was adopted with <5% error.

Determining the vulnerability of the population due to crater formation was straightforward. People located within the final crater zone at the time of impact had no chance of survival and, thus, vulnerability was unity in this area. On the other hand, people outside the final crater zone were not affected by cratering. In this research, world grid data were employed with a cell resolution of about $4.6 \times 4.6$ km². Cratering vulnerability in a given grid cell was determined by calculating the fraction of the crater area that covers this specific grid cell with respect to the grid cell area. Note that the impact point grid cell might be covered completely by the crater but that cells that are located on the rim of the crater are only partially covered and ARMOR's algorithm accounts for such situations. To this end, the final crater area $F_{fr}$ was assumed to be circular:

$$F_{fr} = \pi \left(\frac{D_{fr}}{2}\right)^2 \qquad (26)$$

**Seismic Shaking**

The seismic shock is expressed in terms of the Gutenberg-Richter scale magnitude. It is assumed that a fraction of $10^{-4}$ of the impacting kinetic energy is transformed into seismic shaking (Schultz and Gault 1975). The Gutenberg-Richter magnitude energy relation provided the magnitude of the expected shock as:

$$M = 0.67 \log_{10} E - 5.87 \qquad (27)$$

where $E$ is the impacting kinetic energy in Joules, and $M$ is the magnitude on the Richter scale. With increasing distance from the impact site, the force of the shocks decreases and (Collins et al. 2005) present an empirical law that describes the effective magnitude $M_{eff}$ at a distance $D$ from the impact site:

$$M_{eff} = \begin{cases} M - 2.38 \times 10^{-5} D & | \quad D < 60\ km \\ M - 4.8 \times 10^{-6} D - 1.1644 & | \quad 60 \leq D < 700 km \\ M - 1.66 \log_{10} \Delta - 6.399 & | \quad 700 km \leq D \end{cases} \qquad (28)$$

It should be noted that seismic shaking severity estimation (as is true for other effect severity estimation) carries uncertainty in itself and could produce casualty estimation errors beyond what is captured in the best and worst cases for the vulnerability models.

The effective magnitude can be related to the expected destruction at the given distance from the impact point and determines vulnerability.

A literature review was conducted to find suitable data to support a seismic vulnerability model. Specifically, data were needed to relate seismic shaking magnitude at a given location to the mortality rate at this location. However, typical earthquake records report only peak intensity and total losses and these data are too convoluted for usage here because they depend on population density and affected area in the location of the earthquake which are typically not reported in the respective publications (Norlund et al. 2009). The data reports fatalities that occur in an area that encompasses the entire earthquake region and relates this casualty figure to the peak intensity shaking. However, not all fatalities occur at the location of peak shaking intensity (the epicentre) and some casualties are found at a distance away from the epicentre. Thus, it would be wrong to use these data because they attribute the casualties of the entire earthquake region to the peak shaking intensity and would produce an overestimation for a given seismic intensity.

A function is needed that provides the mortality rate with respect to local shaking magnitude because mortality varies with distance from the epicentre. In other words, an earthquake produces a high mortality rate close to the epicentre where seismic shaking is severe and a lower mortality rate at a distance from the epicentre because seismic shaking attenuates with greater distance. In (Wu et al. 2015), mortality rates are provided as logistic functions with respect to seismic intensity based on earthquake records in China and these functions were validated against four severe earthquake events. It should be noted that the reported mortality rates are equivalent to the vulnerability rates that are of interest here because the mortality rates describe the observed number of casualties for a given seismic shaking intensity. The vulnerability logistic function that best fitted the validation data (mean estimation error of 12%) with seismic intensity (Modified Mercalli Intensity (US Geological Survey 2015)) as free parameter is:

$$V_{i_{seis}} = \frac{1}{0.01 + 2.691 \times 10^6 \times 0.170^{i_{seis}}} \quad (29)$$

In table 2 of (Collins et al. 2005), the necessary data to translate Modified Mercalli Intensity into Richter scale magnitude values is provided. Here, a linear function was least square fitted to the data ($R^2 = 0.9887$) and it is:

$$i_{seis} = 1.4199 M_{eff} - 1.3787 \quad (30)$$

Equations (29) and (30) can be combined into a new sigmoid function that yields expected vulnerability with effective shaking expressed in Richter scale magnitude as free parameter:

$$V_{seis}^{case} = a \frac{1}{1 + e^{b(M_{eff}+c)}} \quad (31)$$

Furthermore, Figure 4 as well as Table 5 of (Wu et al. 2015) supplies data about the variability in vulnerability estimates. Based on this additional information the curves for best and worst case vulnerability to seismic shaking were established, and the corresponding coefficients are:

Table 5: Seismic shaking vulnerability coefficients.

| Case | a | b | c |
|---|---|---|---|
| Expected | 1.0 | $-2.516 \times 10^0$ | $-8.686 \times 10^0$ |
| Best | 1.0 | $-2.508 \times 10^0$ | $-9.590 \times 10^0$ |
| Worst | 1.0 | $-3.797 \times 10^0$ | $-7.600 \times 10^0$ |

Figure 11 shows the seismic vulnerability functions over the expected range of seismic shaking magnitudes.

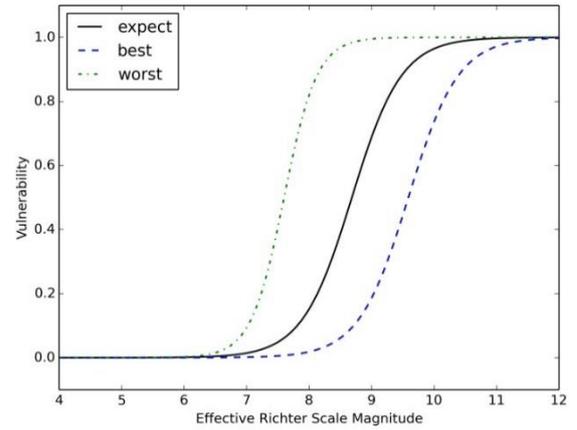

Figure 11: Seismic shaking vulnerability models as a function of effective Richter scale magnitude.

**Ejecta Blanket Deposition**

In addition to plastically deforming and partially melting the impact site, the asteroid impact also ejects ground material outwards from its impact site and the removed material is called ejecta. Ejecta blanket deposition was modelled, and this hazard can lead to delayed damage such as building collapse due to the accumulating ejecta load on structures. The description of this phenomenon has an empirical basis in the literature based on experience with volcanic ash deposition (Pomonis et al. 1999). Reference (Collins et al. 2005) derives an analytical expression for ejecta blanket thickness $t_e$ as a function of transient crater size $D_{tc}$ and distance from impact site $D$:

$$t_e = \frac{D_{tc}^4}{112 D^3} \quad (32)$$

Ejecta deposition is a hazard because it can lead to the collapse of buildings if the weight load of the settling ejecta blanket becomes large enough. The vulnerability model used in this work follows closely the method described in (Norlund 2013) and a mean ejecta material density of $\rho_e = 1600$ kg/m³ is assumed. Given the ejecta density $\rho_e$, ejecta blanket thickness $t_e$ and the standard gravitational acceleration $g_0$, the load of the ejecta blanket is:

$$p_e = t_e \rho_e g_0 \qquad (33)$$

In (Pomonis et al. 1999), it is estimated that 20% of the occupants in a house would be trapped in the event of a collapse and half of those would be fatalities. Keeping with previous assumptions that 22% of the population would be outside at any given time, the remaining 78% are located indoors. Taking these factors together, the maximum vulnerability of the population in the event of a roof collapse is $0.78 \times 0.2 \times 0.5 = 0.078$. However, to realize this vulnerability the roof of a building has to first collapse. The likelihood of roof collapse can be modelled as a function of ejecta load as well as building strength and the corresponding models were derived in (Pomonis et al. 1999). The resulting vulnerability model for the expected case that assumes medium strength housing is:

$$V_e^{expect} = 0.078 \times \left[1 + e^{-1.37(p_e - 3.14)}\right]^{-4.6} \qquad (34)$$

In the best and worst case models, strong and weak building strengths were assumed, respectively. The corresponding vulnerability models are:

$$V_e^{best} = 0.078 \times \left[1 + e^{-1.00(p_e - 5.84)}\right]^{-2.58} \qquad (35)$$

$$V_e^{worst} = 0.078 \times \left[1 + e^{-4.32(p_e - 1.61)}\right]^{-4.13} \qquad (36)$$

Figure 12 visualizes the vulnerability models as a function of ejecta blanket thickness.

In addition to loading of buildings by ejecta material, structural integrity may also be weakened through the impacts of large and fast ejecta projectiles. This aggravating effect was not considered here.

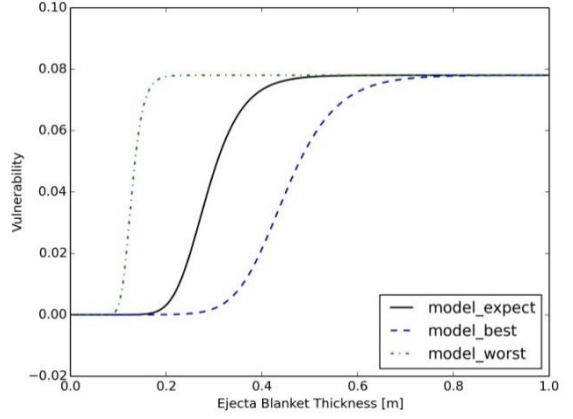

Figure 12: Ejecta blanket thickness vulnerability models.

**Tsunami**

For asteroids that reach the Earth surface intact, a water impact is twice as likely as a ground impact because water covers about double the surface area relative to ground and asteroid impacts show a near-uniform distribution globally (Rumpf et al. 2016b). When an asteroid impacts a water surface, a circular wave pattern is generated and these waves may reach tens of meters in amplitude. Such waves are referred to as tsunamis and, when large enough, they cause devastation at coastlines where they inundate inhabited zones.

Tsunami modelling due to asteroid impacts has received some attention in recent decades but large uncertainties with respect to the expected wave heights, not just in deep water but especially at the interface between sea and land, remain (Wünnemann et al. 2010; Korycansky and Lynett 2007; Ward and Asphaug 2000; Van Dorn et al. 1968; Gisler et al. 2011). However, a comparative analysis was conducted in (Wünnemann et al. 2010) and a suitable, analytical description of wave amplitude $A(D)$ propagation in deep water as a function of transient water cavity diameter $D_{tc}$ and distance $D$ is given as:

$$A(D) = \min(0.14\, D_{tc}, h_{sea}) \frac{D_{tc}}{2D} \qquad (37)$$

where $h_{sea}$ is the ocean depth at the impact site. In accordance with (Korycansky and Lynett 2007), deep water has been defined as any depth >800 m. This equation adopts a $1/D$ wave distance attenuation relationship which seems to match observations made for the 2004 Sumatra tsunami (Weiss et al. 2006; Fritz et al. 2007). The transient cavity diameter $D_{tc}$ may be calculated with equation (24) using the constant factor 1.365 instead of 1.161 (Collins et al. 2005). These relationships were used in ARMOR to propagate the initial wave height through deep water to the point where the bathymetry profile rose to less than 800 m water depth. If the impact point was already located in shallow water, equation (37) was used to calculate initial wave height outside the transient crater for subsequent run-up estimation outlined below.

To determine the coastline affected by an impact generated tsunami, a fast ray tracing algorithm was employed which connects each coastline point inside the reach of the tsunami wave (maximum reach set to roughly 13,000 km) with the impact point. The algorithm traces the coastline map along the connecting rays to detect if an island or another coastline does not obstruct the targeted coastline point. The algorithm has a priori knowledge of how far the closest coastline is away from the impact point and only starts the computationally intensive ray tracing from this distance outwards until it intersects a coastline. Consequently, the algorithm is able to extract such coastlines that are directly "visible" from the impact site and also the corresponding points along the connecting rays where deep water transitions to shallow water (800 m).

Actual wave propagation across the ocean is more complex because refraction of waves around land features, and subsequent interference of the wave with itself may occur. However, the aim of this work was to provide fast models, which did not allow for computationally expensive numerical methods and an analytical solution is presented instead.

The magnification of a tsunami wave as it approaches the shore is called run-up and this parameter ultimately determines how much of the coastal area is inundated and, thus, threatened. Much uncertainty in the field of tsunami modelling arises around if tsunami waves break at the continental shelf in what is coined the "Van Dorn" effect (Van Dorn et al. 1968). If the Van Dorn effect is real, much of the tsunami energy is dissipated at the continental shelf and the run-up of the wave would be greatly diminished (Melosh 2003). To address this issue, run-up is treated in (Korycansky and Lynett 2007) utilizing the concept of the "Irribaren number" $\xi$. The Irribaren number concept is promising because it shows good agreement with various modelling conditions, such as varying bathymetry profiles and wave heights as demonstrated by comparison with numerical results and wave-tank experiments (Korycansky and Lynett 2007). The Irribaren number is defined as (Hunt 1959; Battjes 1974):

$$\xi = s \left(\frac{2A_{800}}{w}\right)^{-\frac{1}{2}} \quad (38)$$

where $s$ is bathymetry slope from deep water to the coast and $w$ is wavelength. The wave amplitude at a sea depth of 800 m is $A_{800}$ and this value is obtained using Equation (37). The Irribaren number approach suggests that run-up $U$ scales as the product of wave amplitude $A_{800}$ and Irribaren number $\xi$:

$$U = 2 A_{800} \xi \quad (39)$$

In the numerical asteroid ocean impact analysis (Gisler et al. 2011) the approximate, empirical relationship, that wavelength, $w$, is double the transient crater diameter, $D_{tc}$, can be found:

$$w = 2D_{tc} \quad (40)$$

Combining equations (38), (39) and (40) yields:

$$U = 2sA_{800}\left(\frac{A_{800}}{D_{tc}}\right)^{-\frac{1}{2}} \quad (41)$$

Run-up calculation requires slope estimation of the bathymetry profile leading up to the exposed shoreline. Suitable, global bathymetry (or, more correctly, hypsography when it also considers land topography) data are available from (Patterson and US National Park Service 2015). Slope is calculated using the simple "rise over run" definition given by:

$$s = \frac{|h_{800}|}{D_{h_{800}-shore}} \quad (42)$$

where $h_{800}$ is the sea depth where deep water stops and $D_{h_{800}-shore}$ is the distance between the $h_{800}$ point and the exposed shoreline in the direction of the approaching wave. Note that a shallower depth than 800 m was used in case the impact occurs in shallow water. The calculated run-up height was used to estimate population vulnerability and inundation range.

*Vulnerability Model*

A tsunami can inundate coastal regions and, thus, harm the population living there. How much of the coastline is inundated depends on the run-up height $U$ and terrain slope $s$. A steeper terrain slope limits the extent to which water of a given run-up height can reach inland. Figure 13 visualizes this concept and in this figure, the red portion of one pixel (pixel length is $p_{len}$) is inundated by a wave with run-up $U$ while the green portion is beyond the reach of the wave because of terrain elevation $\Delta h$ over the length of one pixel.

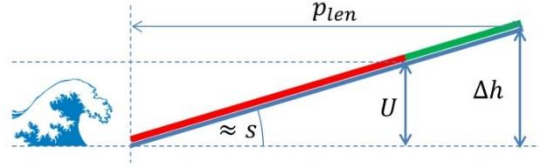

Figure 13: Tsunami run-in as it relates to run-up U and beach slope s. Red terrain in one pixel is inundated and green terrain is safe. Note, that tsunami wave height and run-up are not the same.

Assuming a sufficiently large wave, more than one map pixel may be inundated. In contrast to the linear model portrayed in Figure 13, multi-pixel inundation cannot be calculated linearly, because, depending on the local topography, the terrain slope changes from pixel to pixel. ARMOR's code accounts for this fact by recalculating slope between map pixels and comparing the topography height with the run-up height to determine wave run-in and local pixel inundation. Local inundation is equivalent with local run-up. Further, the code calculates the mean run-up height $U_p$ for each map pixel, whether fully or partially inundated, for subsequent vulnerability estimation. In other words, if $U > \Delta h$, then the entire pixel is inundated and mean run-up in this pixel is $U_p = U - \Delta h_p/2$. In the next map pixel, run-up is reduced by $\Delta h$ of the previous pixel and the new local medium run-up is, thus, $U_p = (U - \Delta h_{p-1}) - \Delta h_p/2$. This formulation can further be expressed as:

$$U_p = U - \sum_{i=0}^{p-1}\Delta h_i - \frac{\Delta h_p}{2} \quad (43)$$

where $i$ counts the pixels from the shore (pixel 1) to the local pixel $p$ and $\Delta h_0 = 0$.

The above equation holds true for completely inundated pixels. The last pixel will generally only be partially inundated, as shown in Figure 13, and, thus, the local run-up is equal to the original run-up less the terrain height of the previous pixels. Finally, the mean local run-up of the last pixel is half of the run-up height at the last pixel:

$$U_p = \frac{U - \sum_{i=1}^{p-1} \Delta h_i}{2} \qquad (44)$$

Furthermore, pixel vulnerability is scaled to pixel exposure - A pixel that is only 10% exposed cannot reach a larger vulnerability than 0.1.

Tsunami mortality as a function of run-up is shown in (Berryman 2005) and this model is based on literature research of tsunami records. The sole parameter in this mortality function is local medium run-up calculated above.

A sigmoid function was fitted to the data provided in (Berryman 2005). Subsequently, the covariance values for the fit were used to establish the best and worst case functions using a variation of $\pm 1\sigma$.

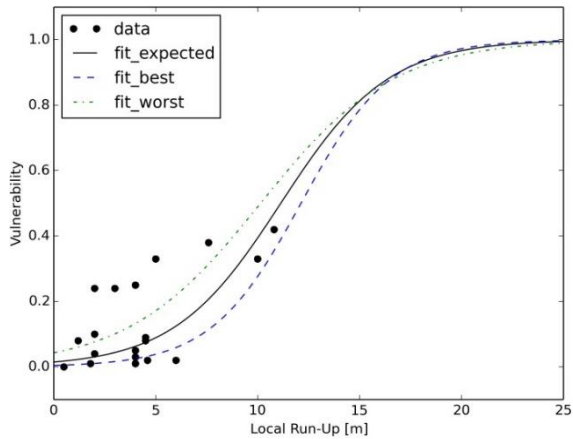

Figure 14: Tsunami vulnerability models as functions of local run-up height with data points from (Berryman 2005).

Figure 14 shows the vulnerability for expected, worst and best outcome according to:

$$V_{\text{tsuna}}^{case}(U_p) = a \frac{1}{1 + e^{b(U_p + c)}} \qquad (45)$$

with the case parameters in Table 6:

Table 6: Tsunami vulnerability model case coefficients.

| Case | a | b | c |
|---|---|---|---|
| Expected | 1.0 | $-3.797 \times 10^{-1}$ | $-1.114 \times 10^{1}$ |
| Best | 1.0 | $-4.528 \times 10^{-1}$ | $-1.213 \times 10^{1}$ |
| Worst | 1.0 | $-3.067 \times 10^{-1}$ | $-1.015 \times 10^{1}$ |

## RESULTS

The ARMOR tool was utilized to estimate total casualties and damage contributions in two case studies as predicted by the impact effect (correct implementation of impact effect models has been demonstrated in (Rumpf et al. 2016a)) and the, here presented, vulnerability models under consideration of the local population and geography.

The first case study was comprised of airburst (50 m sized object) and crater forming (200 m sized object) events over Berlin (Latitude: 52.51°N, Longitude: 13.40°E) and London (Latitude: 51.50°N, Longitude: 0.10°W). For these scenarios an impact speed of 20 km/s, an impact angle of 45° and an asteroid density of 3100 kg/m³ were selected. The results are presented in Table 7 which shows the contribution of each impact effect to the total number of casualties (bold) as well as how the selection of best and worst case vulnerability models affect the total casualty outcome. The damage contributions are colour coded to help discern the most harmful effects.

Table 7: Casualty estimates for airburst (50 m) and cratering (200 m) events over Berlin and London. Impact angle is 45°, impact speed is 20 km/s and asteroid density is 3100 kg/m$^3$.

|  | Berlin | | London | |
|---|---|---|---|---|
| Size [m] | 50 | 200 | 50 | 200 |
| Type | Airburst | Cratering | Airburst | Cratering |
| Wind [%] | 85.5 | 48.6 | 84.6 | 49.1 |
| Pressure [%] | 0.0 | 24.9 | 0.0 | 23.4 |
| Thermal [%] | 14.5 | 23.6 | 15.4 | 24.3 |
| Seismic [%] | 0.0 | 0.1 | 0.0 | 0.1 |
| Cratering [%] | 0.0 | 0.4 | 0.0 | 0.8 |
| Ejecta [%] | 0.0 | 2.4 | 0.0 | 2.2 |
| Tsunami [%] | 0.0 | 0.0 | 0.0 | 0.0 |
| Total expected [-] | 1,180,450 | 3,511,397 | 2,818,507 | 8,761,812 |
| Best [% variation] | -34.1 | -4.0 | -31.6 | -6.2 |
| Worst [% variation] | 35.5 | 4.8 | 32.3 | 7.7 |

The second case study analysed the offshore region of Rio de Janeiro. A 200 m sized object was simulated to impact onshore (Latitude: 22.98°S, Longitude: 43.22°W) and at eight distances offshore moving directly South from the onshore location. The same impact parameters were used as in case study one. The total casualty estimates as well as the damage distributions per impact effect are provided in Table 8. Similarly to Table 7, the damage contributions are colour coded to help interpretation of the results. In addition, the water depth at the impact locations is given. Rio de Janeiro was chosen because it is exposed to tsunamis and because the bathymetry profile is relatively benign with a continental shelf reaching about 120 km offshore and a subsequent transition into deep ocean.

Table 8: Impact scenarios near Rio de Janeiro providing casualty estimates (expected) as well as damage contribution per impact effect in terms of casualty numbers and percentage of total casualties (percentages are colour coded). Asteroid size is 200 m, impact angle is 45°, impact speed is 20 km/s and asteroid density is 3100 kg/m³.

| Offshore [km]   | 0         | 10        | 40      | 100    | 120    | 125    | 130    | 150    | 300    |
|---|---|---|---|---|---|---|---|---|---|
| Water Depth [m] | 0         | 5         | 101     | 166    | 361    | 479    | 580    | 1050   | 2140   |
| Wind [%]        | 55.3      | 59.0      | 82.9    | 91.6   | 55.7   | 40.8   | 24.9   | 0.0    | 0.0    |
| Wind [-]        | 4,204,736 | 1,387,076 | 100,032 | 27,481 | 20,671 | 15,954 | 12,465 | 3      | 0      |
| Pressure [%]    | 12.6      | 5.6       | 0.0     | 0.0    | 0.0    | 0.0    | 0.0    | 0.0    | 0.0    |
| Pressure [-]    | 954,932   | 130,485   | 0       | 0      | 0      | 0      | 0      | 0      | 0      |
| Thermal [%]     | 29.9      | 35.3      | 16.1    | 0.0    | 0.0    | 0.0    | 0.0    | 0.0    | 0.0    |
| Thermal [-]     | 2,272,085 | 830,820   | 19,400  | 0      | 0      | 0      | 0      | 0      | 0      |
| Seismic [%]     | 0.1       | 0.0       | 0.0     | 0.0    | 0.0    | 0.0    | 0.0    | 0.0    | 0.0    |
| Seismic [-]     | 7,450     | 0         | 0       | 0      | 0      | 0      | 0      | 0      | 0      |
| Cratering [%]   | 0.9       | 0.0       | 0.0     | 0.0    | 0.0    | 0.0    | 0.0    | 0.0    | 0.0    |
| Cratering [-]   | 66,227    | 0         | 0       | 0      | 0      | 0      | 0      | 0      | 0      |
| Ejecta [%]      | 1.4       | 0.0       | 0.0     | 0.0    | 0.0    | 0.0    | 0.0    | 0.0    | 0.0    |
| Ejecta [-]      | 103,092   | 0         | 0       | 0      | 0      | 0      | 0      | 0      | 0      |
| Tsunami [%]     | 0.0       | 0.1       | 1.0     | 8.4    | 44.3   | 59.2   | 75.1   | 100.0  | 100.0  |
| Tsunami [-]     | 0         | 2,618     | 1,185   | 2,517  | 16,469 | 23,158 | 37,662 | 26,725 | 11,178 |
| Total [-]       | 7,608,522 | 2,350,999 | 120,617 | 29,998 | 37,140 | 39,112 | 50,127 | 26,728 | 11,178 |

## DISCUSSION

The results of the first case study (Table 7) show that large portions of the populations in the Berlin and London greater area (metropolitan population of 6.0 and 13.9 million people, respectively), were casualties in the simulated airburst and cratering events. It should be noted that a more harmful 200 m impactor is expected to impact only once in every 40,000 years, while a smaller 50 m impactor shows an impact interval of roughly 850 years and might, thus, account for a greater threat overall (Boslough 2013). Aerothermal (wind, pressure, and thermal radiation) effects were most harmful, while ground related effects (seismic shaking, cratering and ejecta) account for only about 3% of the losses in the surface impact event of the 200 m object. The primary reasons for this outcome are the varying effect ranges and generated effect severity. Cratering is absolutely lethal for populations living in the cratering zone but the extend of the crater is very limited compared to far propagating effects. Seismic shaking and ejecta deposition may propagate similarly far as the aerodynamic effects but cause less harm because of lower vulnerability levels associated with their respective severity. While pressure and wind generally act in concert, the harming mechanism is different (pressure causing internal organ injury, wind causing external trauma) and the lethal severity threshold for pressure is larger than for wind. This is why no pressure related casualties were found in the airburst event, but a significant number was reported for the more energetic cratering event. In addition to causing harm to internal organs, a pressure shock could cause structures to collapse which would result in secondary casualties potentially influencing effect casualty contribution percentages as well as overall casualty numbers, and this indirect mechanism has not been considered for pressure effects (but it has been for wind which occurs together with pressure). Thermal radiation causes significant harm and exhibits a tendency to contribute larger proportions of the total casualties for higher energetic events. The relative contribution of aerodynamic and thermal effects for the airbursts agree well with the observations made after the comparatively sized Tunguska event in 1908. There, over 2,000 km$^2$ of forest were flattened due to aerodynamic forces and an area of 300 km$^2$ was charred (Boslough and Crawford 2008; Nemtchinov et al. 1994). The two airburst events showed higher variability in terms of total casualty outcomes than the corresponding surface impacting events. Airburst events are generally less severe than surface impacts and this places airbursts mainly in the transitioning regime (where vulnerability function slope is steepest) of the vulnerability models. Best and worst case vulnerability models show highest variability in this regime and this fact is reflected in the variability of the results.

The second case study (Table 8) focused on the contribution of tsunamis. As reference, a cratering land impact was produced at the shore and the impact location was subsequently moved offshore resulting in varying impact effect contributions. The land and near-coastal impactors (at 10 km) showed a similar damage distribution as the land impactors in case study one. However, as the impact point recedes farther from the inhabited land, the varying effect ranges become apparent. As already observed in case study one, the vulnerability to pressure decreases quicker relative to wind vulnerability. Similarly, thermal radiation has a shorter reach compared to the wind blast which only loses lethality at a distance of 130-150 km. Perhaps surprising is the small contribution of tsunamis to loss in the near-coastal region.

ARMOR's tsunami code is sensitive to sea depth for initial wave height calculation and the coastal region of South America's East exhibits a continental shelf that reaches about 120 km offshore close to Rio de Janeiro. A continental shelf typically features a gentle, constant bathymetry slope $s$ of about 0.1° ($\approx 0.0017$ rad) (The Editors of Encyclopædia Britannica 2016) such that the sea depth $h_{Sea}$ at a given distance from the shore $D$ can be calculated as:

$$h_{sea} = sD \qquad (46)$$

where $s$ is given in radians. Substituting this relationship into equation (37) yields that the coastal height of a wave which is generated by an impact on the shallow continental shelf is independent of distance to the shore and only depends on initial impact energy (reflected in transient crater diameter $D_{tc}$) and slope:

$$A = s\frac{D_{tc}}{2} \qquad (47)$$

For a 200 m impactor such as was utilized here, a transient water cavity of about 5 km in diameter was produced resulting in a wave height of only 4.3 m at the shore according to equation (47). Impacts elsewhere on the shelf will produce similar wave heights at the shore as long as the assumptions about constant slope and shallow water impacts (defined by $h_{sea} < 0.14\, D_{tc}$ in equation (37) corresponding to about 750m for a 5km transient water cavity) hold true. The underlying mechanism is that while sea depth increases seawards, which allows for the generation of larger waves, these waves have to travel farther to the shore attenuating with distance and these opposing mechanisms result in a constant wave height at the shore. Consequently, the continental shelf, through its shallow water and gentle, constant slope, serves a protective function which limits the tsunami wave height that can reach the shore from impacts that occur anywhere on the shelf. This protective effect is different from the Van Dorn effect which describes wave breaking at the edge of the continental shelf (Korycansky and Lynett 2005; Van Dorn et al. 1968). It should be noted that ARMOR's code calculates wave height at the shore according to the Irribaren approach which also accounts for run-up wave height due to shoaling. However, the analytical example (using equation (37)) illustrated more suitably the underlying mechanisms of wave height attenuation over distance and sea depth limited initial wave height to explain the relatively small contribution of tsunami damage in Table 8 at distances corresponding to the continental shelf. The relative damage contribution of tsunamis only started to increase significantly as the impacts reached the edge of the continental shelf and the correspondingly deeper water. This was the point where aerodynamic and tsunami lethality reached equal magnitude (at about 120-125 km) and tsunamis became most harmful in deeper sea. Notably, the possibility for larger waves in deeper water even allowed for an increase in total casualties for impacts at distances of 120-130 km from the shore despite the continuous weakening of aerodynamic effects at those distances.

The results in Table 8 illustrate the significantly farther reach of tsunamis compared to those of other impact effects. Casualty numbers for an impactor at 100 km distance to the shore were 99.6% smaller compared to those produced by a shore impactor because most effects (except tsunami) did not reach the shore from this distance. By comparison, casualty numbers only decreased by about 63% over the next 200 km (between 100 km and 300 km) where tsunamis were the only relevant hazard. The results illustrate why tsunamis pose a significant threat by deep water impacts through their far reach.

While tsunamis generally affect far longer coastlines than could be covered by the other impact effects (especially aerothermal), they usually do not penetrate land inwards as far as other effects if the impact occurs suitably close to the coastline. By reaching farther land inwards, aerothermal effects are, thus, able to cover larger population numbers overall. In addition to the limited initial wave height due to shallow sea depth for near coastal impactors, this observation contributes to the finding of low tsunami casualty numbers relative to those of aerothermal effects.

## CONCLUSIONS

This work derives and presents new vulnerability models that can be used by the wider community to assess the asteroid impact hazard in relation to human populations.

The vulnerability models facilitate the calculation of asteroid impact consequences in terms of loss of human life under consideration of the environmental effects produced in an impact event. The algorithm that is employed by ARMOR to calculate whether an asteroid experiences an airburst or surface impact was presented. Additionally, the algorithm determines effect severity for seven impact effects: strong winds, overpressure shockwave, thermal radiation, seismic shaking, ejecta deposition, cratering and tsunamis. A comprehensive, analytical model for tsunami propagation was presented that is amenable to varying bathymetry profiles. To enable casualty calculation, vulnerability models for all impact effects were derived that connect effect severity to lethality for human populations.

Two case studies examined casualty estimation utilizing the new vulnerability models. In these case studies, it was found that aerothermal impact effects were most harmful with the exception of deep water impacts, where tsunamis became the dominant hazard. Analysing casualty outcomes for ocean impactors in the Rio de Janeiro area revealed the protective function of continental shelfs against the danger of tsunamis by impactors on the shelf. It also illustrated that aerothermal effects, through their farther reach land inwards, can harm a larger population than tsunamis in near-coastal impacts even though tsunamis may cover longer coastlines.

Furthermore, the case studies found seismic shaking, cratering and ejecta deposition to contribute only negligibly to overall casualty numbers.

Casualty numbers of an on-shore impact decreased significantly when the impact location moved few tens of kilometres off-shore due to the fast effect severity attenuation of the dominant aerothermal impact effects with distance.

The calculation of impact consequences facilitates asteroid risk estimation and the concept of risk as well as its applicability to the asteroid impact hazard were introduced. The results produced in the two case studies represent casualty estimates for a given impact threat and location. Casualty estimates correspond to the product of exposure $\psi$ and vulnerability $V(S)$ in risk equation (1). Consequently, the new vulnerability models enable asteroid risk estimation when the impact probability $P$ (third factor in Equation (1)) is taken into account for example in the form of the spatial impact probability distribution (Rumpf et al. 2016b).

## ACKNOWLEDGMENTS

The work is supported by the Marie Curie Initial Training Network Stardust, FP7-PEOPLE-2012-ITN, Grant Agreement 317185. The authors acknowledge the use of the IRIDIS High Performance Computing Facility at the University of Southampton. James G. LaDue from the National Oceanic and Atmospheric Administration (NOAA) supported this work with advice on tornado vulnerability. We extend our sincere thanks to Gareth Collins who has served as thorough reviewer for this paper.